\documentclass[12pt,preprint]{aastex}
\usepackage{graphics}
\usepackage{epsfig,color,longtable,natbib}
\usepackage{latexsym}

\newcommand{\fdust}{\,f$_{\rm d}$ }

\newcommand{\um}{{\,$\mu$m}}
\newcommand{\ums}{{\,$\mu$m} }

\shorttitle{Age Dependence of the Vega Phenomenon}
\shortauthors{Decin et al.}

\begin{document}

\title{AGE DEPENDENCE OF THE VEGA PHENOMENON: OBSERVATIONS}
\author{G.\,Decin}
\affil{Instituut voor Sterrenkunde, KULeuven, Celestijnenlaan
200B, B-3001 Leuven, Belgium} \email{greet@ster.kuleuven.ac.be}
\author{C.\,Dominik}
\affil{Sterrenkundig Instituut `Anton Pannekoek', Kruislaan 403,
NL-1098 SJ Amsterdam, The Netherlands}
\email{dominik@astro.uva.nl}
\author{L.B.F.M.\,Waters}
\affil{Insituut voor Sterrenkunde, KULeuven, Celestijnenlaan 200B,
B-3001 Leuven, Belgium and Sterrenkundig Instituut `Anton
Pannekoek', Kruislaan 403, NL-1098 SJ Amsterdan, The Netherlands}
\email{rensw@astro.uva.nl} \and
\author{C.\,Waelkens}
\affil{Instituut voor Sterrenkunde, KULeuven, Celestijnenlaan
200B, B-3001 Leuven, Belgium}
\email{christoffel@ster.kuleuven.ac.be}

\begin{abstract}
We study the time dependency of Vega-like excesses using infrared 
studies obtained with the imaging
photopolarimeter ISOPHOT on board of ISO. We review the different
studies published on this issue, and critically check and revise
ages and fractional luminosities in the different samples. The
conclusions of our study differ significantly from those obtained
by other authors
\citep[e.g.][]{holland1998Natur.392..788H,spangler2001ApJ...555..932S}
 who suggested that there is a global power-law governing the
 amount of dust seen in debris disks as a function of time. Our
 investigations lead us to conclude that {\sl (i)} for stars at most
 ages, a large spread in fractional luminosity
 occurs, but {\sl (ii)} there are few very young stars with intermediate 
or small excesses;
  {\sl (iii)} the maximum excess seen in stars of a given age is
  about \fdust$\approx 10^{-3}$, independent of time; and {\sl
  (iv)} Vega-like excess is more common in young stars than in old
  stars.
\end{abstract}




\section{INTRODUCTION}
The detection of dust orbiting around Vega
\citep{aumann1984ApJ...278L..23A} and other nearby main-sequence
stars by the IRAS satellite marks the first detection of planetary
material in stellar systems other than our own solar system.  In
the years since then, much attention has been paid to understand
the origin and evolution of this phenomenon, and on the question
if and how this discovery can tell us more about the planetary
systems probably hidden in those dust rings and clouds. One
important aspect of the research was the study of temporal
evolution of the amount of dust present in these systems. Already
\citet{backman1987BAAS...19Q.830B} noticed from colour-colour
studies, that systems like Vega, Fomalhaut and $\epsilon$~Eridani
could be more developed versions (with less dust) of the most
famous Vega-like star \objectname{$\beta$~Pictoris}. This
suspicion was further strengthened by submillimetre observations
of the four main members of this class of objects. Submillimetre
observations are very useful since they have a better chance to
measure the mass present in the system.
\citet{holland1998Natur.392..788H} showed that the measured masses
of the disks of several prominent Vega-like stars seem to follow a
power-law dependence with time. A stunning extra was that even the
zodiacal dust cloud of the solar system seems to fit this trend.

With the launch of the Infrared Space Observatory, ISO, in 1995,
several research groups set out to get a better handle on the time
evolution of the Vega phenomenon. These groups used similar
observations like the ones obtained with IRAS, namely photometry
at mid-IR to far-IR wavelengths to determine the dust masses in
these systems. Several roads where followed to improve the poor
statistical significance of the IRAS studies: IRAS had shown that
some stars display the Vega-phenomenon, but the stars did not form
a good sample in several ways.  First, the ages of the stars were
poorly determined.  And second, the detection limits of IRAS were
such that large excesses could be detected, but for most stars the
detection of the photosphere of the stars was not possible. It was
therefore easier to establish that a star has a dust disk than to
show that it does not have one.

The first issue, concerning the age determination, was addressed
by several groups \citep[e.g.][]{spangler2001ApJ...555..932S} by
choosing stars in different clusters.  Since the ages of clusters
can be determined rather accurately, this strategy should provide
a much better calibration of the time axis.  The disadvantage of
this approach was that most clusters are not very nearby, and the
photosphere of most stars in the sample was out of reach also for
ISO sensitivities.

\citet{habing1999Natur.401..456H,habing2001A&A...365..545H} and
\citet{silverstone2000PhDT........17S} addressed the second
problem of the IRAS results on the presence of a debris disk. They
selected volume-limited samples of stars in order to study the
Vega phenomenon in field stars.  In particular,  they chose
different volumes for stars with different spectral type (i.e.
luminosity) in order to ensure that the photosphere of the star
would be detectable in all cases with good signal-to-noise. This
sample is therefore mainly geared to check whether or not a disk
is present. The obvious problem with field stars is, of course,
that it is much more difficult to determine accurate ages for
these stars, which may make a time-dust mass relation difficult to
discover.

After the launch of ISO, it turned out that the targeted
photometric sensitivities were difficult to reach.  In fact, the
limits could be reached, but only with a changed observational
strategy (mini-maps) which required much more time than foreseen
in the proposals.  This has made the outcome of the studies less
decisive than had been hoped. The task of cleaning up this
question will be one of SIRTF's major goals.  Observers were faced
with the decision to either accept larger limits, or to observe
fewer sources and/or at fewer wavelengths.

Nevertheless, the groups working with ISO came up with age trends
based on their observations.  Studying the disk frequency around
young A stars,
\citet{habing1999Natur.401..456H,habing2001A&A...365..545H}
concluded that there was a very clear break in the presence of a
debris disk around 400~Myr. Most A stars younger than this limit
showed Vega-like IR excesses while most A stars older than this
limit did not show corresponding excesses. However, this work  and
the work of \citet{decin2000A&A...357..533D} also showed that
several stars supposedly much older did also show IR excesses.

\citet{spangler2001ApJ...555..932S} published the results of two
samples: stars in several clusters of various ages, and a number
of field stars with less well determined ages.  In their data,
\citet{spangler2001ApJ...555..932S} could determine a trend in the
time evolution of the dust content of stellar systems. They found
a power-law dependence with a power f$_{\rm d}\propto t^{-1.76}$.
They further argued that a collisional cascade is expected to
produce a $t^{-2}$ dependence, and given the errors in the data
that may well be just what is observed.

However, there are two reasons to doubt the conclusions reached by
\citet{spangler2001ApJ...555..932S}: {\sl (1)} the observed
relation between dust mass and age is not confirmed when large
samples are considered, and {\sl (2)} the expected relation
between age and dust mass as derived by
\citet{spangler2001ApJ...555..932S} omits some important physical
effects. In this paper, we re-analyse observations of IR excess
from Vega stars, and in a forthcoming paper (Dominik \& Decin,
2003, hereafter Paper II) we will reformulate the theory for the
time dependence of dust mass from debris disks. We would like to
argue that the current observational evidence for the occurrence
of this power-law is still very weak.  First, in the cluster
observations, this result is based on very small number
statistics: usually there was only one excess source detected per
cluster (with a maximum of three). Furthermore, in order to arrive at a
representation of the \emph{average} dust content of stars in a
cluster, \citet{spangler2001ApJ...555..932S} average all the
derived excesses. Due to the relatively low S/N in the
observations, they also include in this calculation ``negative''
excesses.  It is clear that such an approach can have strong
problems when small numbers are involved. In particular, if only
one source in each cluster has a clearly detectable excess, the
procedure described above would dilute the measured IR excess. The
average dust mass then depends on the number of observed stars in
a certain cluster. Since this number varies per cluster, strong
systematic effects are to be expected. A better approach would be
to look only at the sources with clearly detected excesses.  If
there is a power-law time dependence intrinsic to these sources,
then it should also be visible by just looking at the strongest
excess stars of each age. Finally,
\citet{spangler2001ApJ...555..932S} included in their plots also
points obtained for the Chameleon I, Scorpius and Taurus
star-forming clouds. These stars are clearly pre-main-sequence
objects with first-generation disks in which the formation of
larger bodies is incipient, while the dust in debris disks is
thought to be replenished by collisions of e.g. comets. Moreover,
the disks in pre-main-sequence stars are often optically thick in
the radial direction, so that the IR luminosity in such sources is
determined by the geometry of the disk (flaring or non flaring,
for example) and not by the dust mass in the disk. These stars
should therefore be excluded when considering the evolution of
second-generation dust.

The purpose of this study is to re-investigate the time dependence
of Vega-like excesses from photometric studies with the ISO
satellite. The paper is organised as follows. A list of stars with
IR excess detected by the ISO satellite is compiled in
Sect.~\ref{sec:sample-select-param} and a critical assessment of
the determinations of age and IR-to-total fractional luminosity is
made. In the following section we plot the fractional luminosity
versus age for various samples, and discuss these diagrams in the
framework of the hypotheses alluded to in this introduction. In
Sect.~\ref{sec:Discussion} we discuss the results and derive
lessons and challenges for the coming SIRTF observations.

\section{PARAMETER DETERMINATION}
\label{sec:sample-select-param}

The data we discuss in this paper were taken from
\citet{habing1999Natur.401..456H,habing2001A&A...365..545H,decin2000A&A...357..533D,silverstone2000PhDT........17S,spangler2001ApJ...555..932S}
and Decin et al. (2003, in preparation). They all concern stars
with an IR excess which is most probable caused by dust in a
stellar disk and were obtained with ISOPHOT, the imaging
photopolarimeter on board of ISO
\citep{klaas1994_iom,kessler1996A&A...315L..27K}. All data in the
samples of
\citet{habing1999Natur.401..456H,habing2001A&A...365..545H,decin2000A&A...357..533D}
are C100 mini-maps at 60~\um, while the samples of
\citet{silverstone2000PhDT........17S,spangler2001ApJ...555..932S}
consist of a mixture of mini-maps and chopped C100 observations.
In order to limit the discussion to debris disks and not to
pollute the sample with first-generation disks, we only consider
stars older than 10~Myr.

\subsection{Age Determination}
A variety of observational tests have been
formulated to determine the age of main-sequence stars, which
are based on different physical phenomena and can be applied to different
stellar types with varying success. The dating of open clusters is
considered as most reliable, e.g., by fitting the cluster member
stars' location on Hertzsprung-Russell diagrams to models of
stellar evolution. This method is used for most stars in the
sample of \citet{spangler2001ApJ...555..932S}. The disadvantage of
this approach for our purposes is that most clusters are far away, so that only
disks with large excesses can be securely identified while small
excesses will be difficult to distinguish from photospheric
fluxes.

The other groups worked with single field stars, having a more
difficult and uncertain age determination
(\citep{decin2000A&A...357..533D, habing2001A&A...365..545H,
silverstone2000PhDT........17S} and some stars of
\citet{spangler2001ApJ...555..932S}). An analysis of the different
methods used for age determination for main-sequence stars was
done by \citet{lachaume1999A&A...348..897L}. They concluded that
isochrones are the best age estimators for stars with spectral
type in the range B9--G5. A reliable age determination depends not
only on the accuracy of the temperature and the luminosity
determination, but also on that of the metallicity. Since stars
evolve faster when they age, accurate age determinations are often
more difficult for stars which are still close to the ZAMS. A far
reaching ambiguity may arise when it is not clear if a star is
evolved or still in the pre-main-sequence phase of evolution. A
famous case is \objectname{$\beta$~Pic}: from isochrones it was
suggested that it may be 280~Myrs old.  More recently, the
discovery of several young M dwarfs which are in a moving group
with \objectname{$\beta$~Pic} has shown that 15~Myr
\citep{barrado1999ApJ...520L.123B} is in fact a more likely age
for this star, as was suggested by
\citet{lanz1995ApJ...447L..41L}.

For late-type stars, \citet{lachaume1999A&A...348..897L} concluded
that a variety of methods (metallicity, rotation, calcium emission
lines, and kinematics) is recommended, with a preference for
age-determination using stellar rotation. However, many stars
require in fact a combination of different methods to properly
bracket their actual age. Beside the example given above of
\objectname{$\beta$~Pic}, also the age of e.g.
\objectname{HD207129} was in dispute for several years. While
\citet{jourdain1999A&A...350..875J} deduced a lifetime of 4.6~Gyr
based on the weakness of the CaII K-line emission;
\citet{zuckerman2000ApJ...535..959Z} found an age estimate of
$\sim$40~Myr relying on space motions and location in space.

All stars under study were subject to the same strategy to obtain
a reliable age determination. Priority is given to ages determined
by the lifetime of the cluster or moving group to which the star
belongs. When it is not known whether a star is member of a moving
group or cluster, and the star is an element of a binary system
for which both stars can be placed on the Hertzsprung-Russell
diagram, the age as determined by isochrone-fitting is taken. When
none of these two conditions are met, the star is not on the
ZAMS, and has a spectral type between B9 and G5, then we prefer
isochrone-fitting to determine the age. Otherwise, the age as
determined from rotation or chromospherical activity is taken.

\subsection{Fractional Luminosity}

The second parameter needed in our relationship is the fractional
dust luminosity,
\begin{equation}
\rm{f_d} \equiv F^{\rm exc}_{\rm bol}/F^{\rm photosp}_{\rm
bol}=L^{\rm dust}/L^{\rm bol}\,\,\,,
\end{equation}
which is a commonly used measure of the amount of dust in these
systems.

The different groups adopted different criteria for considering an
objects as displaying an excess. \citet{decin2000A&A...357..533D}
and \citet{habing2001A&A...365..545H} accepted only measurements
where both the detection itself and the excess are measured with
S/N $\ga$3; here the noise is calculated as the gaussian
dispersion of the excess around zero for the non-excess stars,
leading to a dispersion of $\sim$30~mJy at 60~\um.
\citet{spangler2001ApJ...555..932S} considered a source as
detected if the flux at the on-target (centre) pixel exceeded 3
times the standard deviation of the flux on the 8 surrounding
pixels; they obtained a standard deviation comparable with
\citet{decin2000A&A...357..533D} and
\citet{habing2001A&A...365..545H}; however, an excess is accepted
when it has a S/N $\ga$1. \citet{silverstone2000PhDT........17S}
used the same standard deviation calculation as
\citet{spangler2001ApJ...555..932S}, but used a S/N $\ga$2
criterion for the detection itself and for the excess measurement.
To have a more consistent sample, we have refined the sample of
Spangler by requiring a S/N ratio of the excess larger than 2;
this led to the discarding of four objects.

The determination of the fractional luminosity, $\rm{f_d}$, is in
principle straightforward if the excess radiation emitted by the
dust can be measured at enough wavelengths. However, if the dust
flux is measured at only one or a few points, one needs to make
some assumption about the shape of the dust emission spectrum.
While \citet{decin2000A&A...357..533D} and
\citet{habing2001A&A...365..545H} derived the fractional
luminosity based on the excess measurement at 60~\um, making an
assumption about the temperature of the dust as well,
\citet{spangler2001ApJ...555..932S} and
\citet{silverstone2000PhDT........17S} calculated values of
$\rm{f_d}$ for each of their excess sources following the method
of \citet{backman1987csss....5..340B}, summing the luminosities in
each wavelength band and including a correction to account for
excess flux from wavelengths longer than the 90 or 100~\ums band.
However, for the vast majority of their targets older than
$\sim$20~Myr, there was either no measurable excess emission at 12
and 25~\ums or no flux measurement available. Therefore they
calculated a correction factor necessary to reproduce the final
$f_d$ values from only the 60 and 100~\ums excess, based on stars
for which flux measurements at different wavelengths were known.
Both methods give similar results except for two stars from
Spangler's sample, where their method led to an extremely high
value for the fractional luminosity compared with the other method
and compared to the results for similar stars. For both stars the
fractional luminosity as obtained by the method of
\citet{decin2000A&A...357..533D} and
\citet{habing2001A&A...365..545H} is used during further
discussion in this paper.

\subsection{Overview of the Results}
In this section, we report on the results of the determination of
ages and fractional luminosities for the objects of the five
samples considered. All results are summarised in
Table~\ref{data_disk_evolution}.

\begin{itemize}

\item{\sl{\bf \citet{habing2001A&A...365..545H}}}\\
This sample was selected from the catalogue of stars within 25~pc
from the Sun by \citet{woolley1970ROAn....5.....W}, and contained
those main-sequence stars (spectral type A till K) for which the
photospheric flux at 60~\ums is above the ISOPHOT detection limit
(30~mJy). The sample was also restricted to those stars for which
the infrared flux density can be unequivocally attributed to the
target star. From a list of 84 main-sequence stars, 14 stars were
found having a 3-$\sigma$ infrared excess, i.e. exactly the same
fraction as in the G dwarf sample of
\citet{decin2000A&A...357..533D}. From the final list with stars
with an IR excess, \objectname{55~Cnc} is omitted as Vega-type
star because the sub-millimetre flux, initially attributed to dust
orbiting the star \citep{dominik1998A&A...329L..53D}, is reported
instead to be from nearby background sources
\citep{greaves2000AAS...197.0826G}.

The ages were estimated by \citet{lachaume1999A&A...348..897L}
using isochrones for the early-type stars, and stellar rotation or
a combination of methods for late-type stars. Since, these
determinations have been questioned for two objects of the sample
which appear to belong to moving groups of considerably shorter
age. As already mentioned, \citet{barrado1999ApJ...520L.123B}
associated \objectname{$\beta$~Pic} with several co-moving young M
dwarfs, the age of which they estimated at 15~Myr. From a more
extensive study of this moving group,
\citet{zuckerman2001ApJ...562L..87Z} lowered this age to the
12~Myr we adopt in this study.
\citet{zuckerman2000ApJ...535..959Z} suggested that also
\objectname{HD207129} belongs to a young moving group, the Tucanae
Association, for which they estimate the age at 40~Myr, i.e. two
orders of magnitude shorter than the result of the isochrone and
the CaII study by \citet{lachaume1999A&A...348..897L}. In support
of their shorter age for \objectname{HD207129}, which we also
adopt, \citep{zuckerman2000ApJ...535..959Z} mention, besides the
kinematic evidence, also high lithium abundance and an X-ray
luminosity which is ten times that of the Sun.

\item {\sl{\bf \citet{decin2000A&A...357..533D}}}\\
This sample was selected from the CORALIE planet-search programme,
and consists mainly of G dwarfs \citep{udry2000fepc.conf..571U}.
The CORALIE distance criterion locates the chosen stars in the
Local Bubble, reducing the possibility that the dust responsible
for excess thermal emission originates from the interstellar
medium rather than from a planetary debris system.
\citet{decin2000A&A...357..533D} found 5 excess stars  with a
S/N$\ge$3 out of the 30 measured targets. Since the sample
selection was free from bias with  respect to infrared excess,
this works suggests that some 17\% of G dwarfs possess debris
disks.

Since a fairly conservative approach was used for the selection of
the excess stars, we are confident that the fractional
luminosities were determined reliably.  The ages were estimated
from isochrones and confirmed -- when possible -- by other
methods. There is no indication that any of these stars is
particularly young. Four objects occur rather close to the ZAMS in
the HR-diagram, so that their ages determined from isochrones may
be rather uncertain, but the star \objectname{HD22484} definitely
is well evolved. By all means, the fact that sizeable fraction of
a fairly unbiased sample of G dwarfs displays excesses, strongly
suggests that relatively massive debris disks may survive around
main-sequence stars of several Gyr old.

\item{\sl{\bf \citet{spangler2001ApJ...555..932S}}}\\
The targets in this sample fall in three distinct categories: {\sl
(1)} main-sequence stars in relatively nearby (generally closer
than 120~pc) open clusters including $\alpha$~Persei, Coma
Berenices, Hyades, Pleiades, and the Ursa Major nucleus and
stream. Cluster ages are between 50 and 700~Myr, and the target
stars span spectral types A--K. {\sl (2)} Selected classical and
weak-line T~Tauri stars in the Chameleon~I, Scorpius, and Taurus
star-forming clouds, and {\sl (3)} a small sample of relatively
nearby (less than 60~pc) isolated stars with indications of youth.
As mentioned above, the sample of the T~Tauri stars is rejected
from our study. We want to be sure that the dust is clearly 
`second generation', i.e. not primordial but released from larger bodies
such as asteroids and comets. Four more objects 
(\objectname{HII~3163}, \objectname{HD17796},
\objectname{HD184960} and \objectname{HD27459}) where discarded,
because the S/N of the claimed excess measurement is lower than 2.

As alluded to in the previous section, the fractional luminosity
of \objectname{HII~1132} (Pleiades cluster) and
\objectname{HD72905} (Ursa Major cluster) were adapted to lower
values. In total, there were 3 excess detections in both Coma
Berenices and Ursa Major, 1 detection in both the $\alpha$~Persei
and the Pleiades clusters, and no detections in the Hyades. The
fraction of excess stars discovered in these clusters is smaller
than in the field, probably mainly reflecting the different
sensitivity limits of the surveys. The nearby field stars with an
IR excess considered by \citet{spangler2001ApJ...555..932S} amount
to seven: \objectname{HD105}, \objectname{HD35850},
\objectname{HD37484}, \objectname{HD134319},
\objectname{HD151044}, \objectname{HD202917}, and
\objectname{HD209253}.

Age determinations for cluster objects can be considered to be
robust. For \objectname{HD37484}, the age was determined using the
abundance of lithium as measured by
\citet{favata1993A&A...277..428F} in comparison with the lithium
abundances of stars in the Taurus, the Hyades and the Pleiades
cluster. An extrapolation is however necessary for this star,
which makes the age determination very inaccurate. This star may
in fact be a pre-main-sequence star
\citep{favata1993A&A...277..428F}. For all these reasons, this
star is omitted from the final sample. As dating technique for the
other field stars, \citet{spangler2001ApJ...555..932S} used
literature values of the chromospheric emission as measured in the
cores of the Ca II H and K lines. Since significant additional
information is available for all 6 stars, we have reconsidered
their age determination.

A literature study shows that four of them are members of moving
groups or association: \objectname{HD105} and
\objectname{HD202917} belong to the Local Association
\citep{montes2001MNRAS.328...45M} which is 20-150~Myr old;
\objectname{HD35850} is a member of the $\beta$~Pictoris moving
group \citep{zuckerman2001ApJ...562L..87Z}, which is 12~Myr old.
\objectname{HD134319} is an element of the Hyades supercluster
(600~Myr) \citep{montes2001MNRAS.328...45M}. With respect to the
calcium ages, we then find that \objectname{HD105} and especially
\objectname{HD35850} are much younger, while the difference are
smaller for the two other stars.

Following \citet{lachaume1999A&A...348..897L}, we checked the ages
for the two remaining field stars, which have a spectral type
between B9 and G5,  using isochrones by Claret
\citep{claret1995A&AS..109..441C,claret1995A&AS..114..549C} with
metallicity (X=0.73, Z=0.01), (X=0.70, Z=0.02) and
(X=0.75,Z=0.03). We found that $\log{\rm age[yr]}<8.65$ for
\objectname{HD209253}, which is in good agreement with the age
obtained by \citet{spangler2001ApJ...555..932S}. For
\objectname{HD151044}, the age we found is some 4 times larger
than that quoted by \citet{spangler2001ApJ...555..932S}.

\item{\sl{\bf \citet{silverstone2000PhDT........17S}}}\\
This sample includes four sub-samples. {\sl (1)} The initial focus
was on an unbiased survey of nearby stars, with care taken to span
the stellar parameters of mass, age, and multiplicity. The targets
were selected from the Gliese catalog (a distance-limited catalog
of stars closer than 25~pc). The distance limit was chosen such
that the expected range of fractional luminosity, scaled to the
apparent brightness of the targets, would be brighter than the
anticipated ISO detection limits. Spectral types of the
main-sequence targets were chosen to span the mass range of about
0.5 to 2~M$_\odot$. They preferred stars whose ages were known in
the literature. {\sl (2)} In addition to this distance-limited
survey, a selection of stars with indication of youth ($\log{\rm
age} \sim 8.5$), but whose distances placed them beyond the
distance cutoff described above, was added. {\sl (3)} Because ISO
turned out not to be as sensitive as anticipated, mid-way through
the project the unbiased survey was abandoned in favour of
observations of Vega phenomenon candidates with good IRAS 60~\ums
measurement, but generally only upper limits at 100~\um. These
targets were selected from the IRAS Faint Star Catalog (FSC) file
and the SAO catalog. {\sl (4)} Finally, a selection was made
consisting of 23 stars from the Bright Star Catalog and with
spectral types from A to G, located on the main sequence and
occurring in the region of the sky missed by the IRAS all-sky
survey.

\citet{silverstone2000PhDT........17S} estimated ages from the Ca
II H and K line strengths. We have redetermined the ages for the
stars with a 2-$\sigma$ excess detection and for which the age, as
calculated by \citet{silverstone2000PhDT........17S}, is higher
than 10~Myr. From a literature study, we found that
\objectname{HD25457} is a member of the Local Association
\citep{montes2001MNRAS.328...45M} and that \objectname{HD164249}
belongs to the $\beta$~Pictoris moving group
\citep{zuckerman2001ApJ...562L..87Z}. \objectname{HD4614} and
\objectname{HD131156} are components of nearby visual binary
stars, which when put on isochrones, were found to have ages of
$4\pm 2$~Gyr and $2\pm 2$~Gyr respectively
\citep{fernandes1998A&A...338..455F}. For the remaining stars with
a spectral type between B9 and G5, we have checked the age using
the isochrones of Claret
\citep{claret1995A&AS..109..441C,claret1995A&AS..114..549C} with
metallicity (X=0.73, Z=0.01), (X=0.70, Z=0.02) and
(X=0.75,Z=0.03). The ages obtained by this method give on average
an age which is 0.39~dex larger in $\log{\rm t[yr]}$, which we
consider to be within the errors of the different age
determination methods. For two objects no independent age estimate
could be made, namely the K star \objectname{HD165341} and the F
star \objectname{HD38207} for which no accurate parallax is
available.

The star HD35850 deserves extra attention: it is the only young
star of the whole sample with relatively low fractional
luminosity. The young age of this star appears rather secure: not
only does it belong to the $\beta$-Pictoris moving group, but also
the high lithium content and the high X-ray flux definitely points
towards a genuine youth for this star
\citet{wichmann2003A&A...399..983W}. On the other hand, we feel
that the identification of this star as a Vega-type object may
need confirmation. Looking in detail to the ISOPHOT mini-map, the
observations show -- in contrary to the maps of other Vega-excess
stars -- for only half of the pixels the highest flux when centred
on the star. The HIRAS maps at 60 and 100~\ums of the region
around HD35850 (Fig.~\ref{iras_hd35850}) show an area of extended
cirrus with lot of structure and different equally-bright points.
The rectangle, plotted in the middle of these maps, is an
indication of the size of the region seen by the ISOPHOT mini-map
(not on the true position due to the small mis-pointing of ISO and
IRAS). From inspection of these HIRAS maps, the possibility arises
that the IR excess of HD35850 is related to the star being located
in a zone of extended cirrus.

\item{\sl{\bf The K Giant \objectname{HD3627}, (Decin et al. 2003, in preparation)}}\\
\objectname{HD3627} has a strong IR excess at 60 and 90~\ums and
is an ISOPHOT-C100 point-source (Decin et al. 2003, in
preparation). The excess is most probably due to a debris disk and
not to cirrus or an extended source. \objectname{HD3627} is at the
moment the only known ISOPHOT C100 point-source K giant and is
most probably the successor of an F-type main-sequence star. All K
giants with IR excess from the list of Decin et al. (2003, in
preparation) and \citet{kim2001ApJ...550.1000K} are rejected as
debris-disk stars due to the fact that the IR excess is extended.

\end{itemize}

The parameters for all stars are listed in
Table~\ref{data_disk_evolution}. The first column gives the name
of the stellar target. The 2$^{\rm nd}$, 3$^{\rm th}$ and 5$^{\rm
th}$ column tabulate respectively the V magnitude, the B-V colour
and the spectral type as listed by Simbad. The parallax, as
measured by Hipparcos, can be found in column 4. The logarithm of
the {\sl published} age  as determined by the different mentioned
authors is given in the 6$^{\rm th}$, while column 7$^{\rm th}$
gives the fractional luminosity as we determined it. Column 8
contains the {\sl revised} ages as obtained by the above mentioned
strategy and the reference to these ages are tabulated in column
9. In the last column the relative {\sl reliability} of the
revised age determination is indicated on the basis of the
following criteria: 4: sure age because star is member of a
cluster or moving group; 3: object is member of a binary system
for which both components fit on the same isochrone; 2: age of the
field star could be determined by two methods (isochrones and
chromospherically activity) with a maximal difference of 0.40~dex
in $\log{\rm age}$ (uncertainty for isochrone-fitting is
$\sim$0.40~dex and for chromospherically activity $\sim$0.2~dex
\citep{lachaume1999A&A...348..897L}); 1: large uncertainty on the
age because the age could only be determined using one method, or
two methods give an age difference of more than 0.40~dex in
$\log{\rm age}$. The only exception on these criteria is HD35850.
While the age determination of this star is secure, it is
questionable if the IR excess tentatively detected by ISOPHOT is
connected to the source. For this reason, the relative reliability
of this star is reduced to 1.

\placetable{data_disk_evolution}

\section{RELATIONSHIP BETWEEN LIFETIME AND FRACTIONAL LUMINOSITY}
\label{sec:relation}

\begin{figure}
\plotone{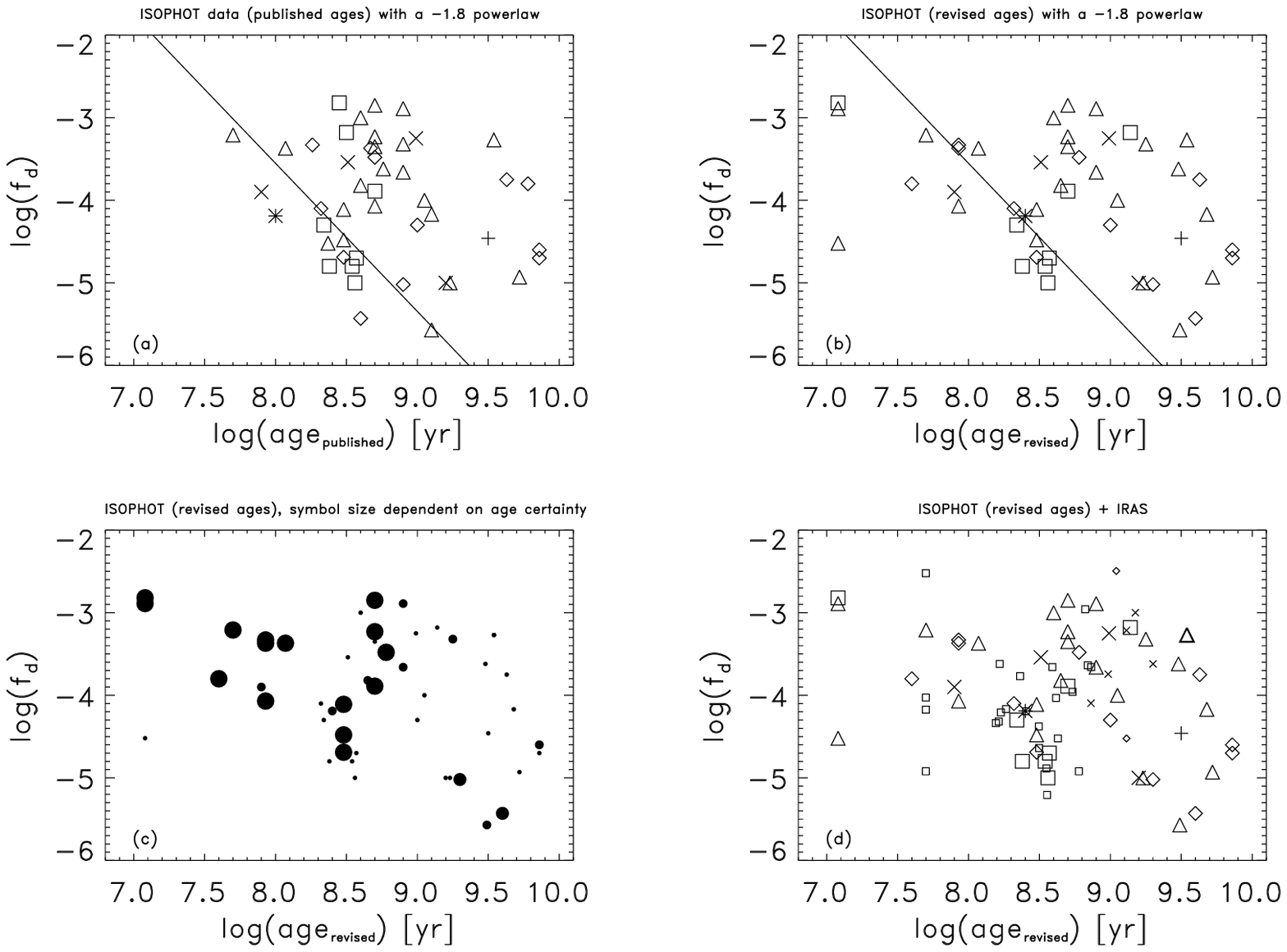}
        \caption{Fractional luminosity of the IR excess as a function of
        age. (a) The data obtained by the different groups using
        the ISOPHOT C100 camera \citep{decin2000A&A...357..533D,habing2001A&A...365..545H,
        silverstone2000PhDT........17S,
        spangler2001ApJ...555..932S}. Explanation of the symbols:
        $\Box$: A main-sequence stars, $\bigtriangleup$: F
        main-sequence stars, $\Diamond$: G dwarfs, $\times$: K
        dwarfs and $+$: K giant HD3627 (Decin et al. 2003, in preparation). (b) The
        {\sl revised} age (see text) versus the fractional
        luminosity. (a) and (b) are over-plotted with the regression line obtained by
        \citet{spangler2001ApJ...555..932S} with slope -1.76.
        (c) Same as (b) but the relative symbol size
        depends on the reliability of the age determination (see
        text). (d) Same as (b) but the sample is increased by the IRAS sample from
        \citet{song2000ApJ...533L..41S,song2001ApJ...546..352S} (small symbols) to
        cheque the ISOHOT data.}
        \label{age_fd}
\end{figure}

The results listed in table~\ref{data_disk_evolution} are
illustrated in Fig~\ref{age_fd}. In Fig.~\ref{age_fd}(a) the
fractional luminosity is plotted as a function of the {\sl
published} age for all samples mentioned above. In
Fig.~\ref{age_fd}(b), the same data is plotted, but now over the
{\sl revised} ages as described in the current paper.
Figures~\ref{age_fd}(a) and \ref{age_fd}(b) also show the
power-law found by \citet{spangler2001ApJ...555..932S}.
Figure~\ref{age_fd}(c) repeats the same data, but now the symbol
sizes reflect the age weight: large symbols show relatively secure
ages, small symbols correspond to large error bars. Finally, in
Fig.~\ref{age_fd}(d) the sample is increased by adding a check-up
IRAS sample obtained from
\citet{song2000ApJ...533L..41S,song2001ApJ...546..352S}. These
data have not been re-evaluated as to their accuracy of age and
fractional luminosity, but they seem to follow the overall
distribution of points derived from ISO observations quite well.

For the following discussion, we will mainly refer to
Fig.~\ref{age_fd}(b), which contains fractional luminosity as
obtained by ISO data versus the revised ages. The distribution of
points in this diagram has a number of features:
\begin{enumerate}
\item\label{item:1} There appears to be a clear cut-off at about
\fdust$\approx 10^{-3}$. The sample contains no Vega-like stars
with significantly higher \fdust-values.
\item\label{item:2} The same upper cut-off of \fdust$\approx 10^{-3}$ is
valid throughout the diagram, and seems to be reached both by
young and old stars. In particular, there are several cluster
stars with well-determined ages between 300 and 700~Myrs which do
have high \fdust-values, far away from the power-law as determined
by \citet{spangler2001ApJ...555..932S}.
\item\label{item:3} The lower-left corner of the diagram is almost
empty, which means there are few young stars with small Vega-like
excesses in the sample. The only example in the current sample is
HD35850, which as we have discussed above, might possibly be the
result of cirrus confusion. With some optimism, one might see a
cut-off line with a slope of $\sim$-1.3, touching to the
lowest \fdust values in each age bin.
\item{\label{item:4}} The lower cut-off for \fdust is about
$10^{-5}$, there are only very few stars below this line.
\end{enumerate}

\section{DISCUSSION}

\label{sec:Discussion}

The different properties for the observed distribution in the
age\,-\fdust diagram, as mentioned above, need a thorough
discussion on their reliability and significance.

\subsection{The upper cut-off: a true limit}
The first important feature of the observed distribution of stars
is the upper cut-off. The conclusion can hardly be escaped that
old stars with fairly high fractional excesses do exist. Though it
is admittedly difficult to obtain precise ages for individual
stars, it is highly unlikely that the fact that the upper right
part of the figure is densely populated, is due to uncertainties
on the ages.

Apparently, there are no stars which physically could be described
as Vega-like stars (following the definition of
\citet{lagrange2000prpl.conf..639G}) with fractional luminosities
above this limit. This result appears to be robust in the sense
that it can not be ascribed to a selection effect. If Vega-type
stars with larger fractional luminosities exist, it is most
unlikely that IRAS would not have detected any of them. The
absence of large excesses must have a physical reason. We will
address this issue in paper II.

\subsection{The lower cut-off: the sensitivity limit}
The lower cut-off in the diagram around \fdust$\approx 10^{-5}$ is
simply the sensitivity limit of the current measurements. The
radiation of smaller amounts of dust is no longer stronger than
the  luminosity of the star at 60~\um, and the dust emission
cannot be detected with significance. Since the dust emission of
Vega-like stars is detected by measuring the difference between
the observed 60~\ums flux and the predicted photospheric flux,
this limit is set by two important factors: the absolute
calibration of the IR photometer, and the accuracy by which we
know the photospheric flux. Therefore, the increased {\sl
sensitivity} of SIRTF alone will not push down this limit
significantly. A much better {\sl absolute calibration} is
necessary as well. Also, significant work on understanding the
photospheres of the individual stars will be required.

\subsection{Few young stars with small amounts of dust?}
The lower left corner of the age\,-\fdust diagram appears to be
empty. Is this feature significant or is it caused by a small
sample size or a selection effect? Table~\ref{data_disk_evolution}
contains 10 stars younger than 200~Myrs. Except for HD35850, all
those stars have excesses above the line touched to the lowest
fractional luminosities in each age bin, with a slope of $\sim$-1.3.
The data are, however, currently not good enough to make a strong
statement about the correctness of the slope. We can only claim
that there is a decline with a slope between -1 and -2. This result
 is also supported by the volume-limited sample
of \citet{habing1999Natur.401..456H,habing2001A&A...365..545H}.
The youngest 6 stars in this sample {\sl all} have a Vega-like
excess. The youngest stars in this sample {\sl without} an excess
is about 200~Myrs old. So, it is quite possible that stars in this
age range and with lower Vega-like excesses are intrinsically
rare. However, it can not yet be excluded that their apparent
scarcity is due to the small number statistics of searches for
such objects as well in the solar neighbourhood and -- for stars
further out -- to the low sensitivity of the observational tools
used so far. SIRTF will certainly provide large enough samples
which can address this question.

\subsection{The age\,-\fdust power-law}
When we want say something about the evolution of debris disks
with time, we have to stress that two effects on this relation
have to be treated separately: {\sl (1)} the dependence of the
dust mass, or correlated \fdust, on age, and {\sl (2)} the
evolution of the fraction of stars displaying the Vega phenomenon.

We see very little evidence for the existence of a power-law
describing globally the dependence of \fdust on age. The errors on
the determination of the age and the fractional luminosity can not
move all Vega-like stars to one single slope in the diagram. Also,
the maximum amount of dust seems to be independent of age. The
decline as found by \citet{spangler2001ApJ...555..932S} relies
heavily on the inclusion in his sample of three very young
clusters for which the infrared excesses stem from proto-stellar
disks; when these clusters are removed, and only genuine Vega-type
systems are considered, the resulting relation is much less clear.
Moreover, we note that \citet{spangler2001ApJ...555..932S} also
detected three fairly large excess stars. These stars are among
others located in Coma Berenices, which is the second oldest
cluster in their sample. These stars can surely not be moved to
the power-law as found by them.

A power-law can, however, be present for individual stars. The
absence of intermediate Vega-like excesses in young stars may be
the strongest evidence for a time evolution of the excess in
individual sources: since at ages of 200~Myrs and up many stars do
have intermediate \fdust values, the absence of younger stars with
similar excesses would mean that stars have evolved from high
\fdust at young ages to low \fdust at 200~Myrs and beyond. As we
will discuss in paper II, this will place interesting conditions
on the structure and mass of young debris disks.

One can also expect that the fraction of stars having a debris
will decrease in time because the debris disks will once be lost.
While clusters should be a good statistical basis to investigate
the real time dependence, the incompleteness (small samples for
which the photosphere is not always detectable) of the sample
today available makes it difficult to address this question. Therefore,
we refer to the volume-limited sample of
\citet{habing1999Natur.401..456H,habing2001A&A...365..545H}. In
this sample, all stars younger than 200~Myr display an IR excess.
The stars which are younger than 400~Myr have 60\% chance to have
a debris disks, while only 9\% of the stars which are older than
400~Myr have a Vega-like IR excess. On average, 16\% of the stars
from this volume-limited sample occurs with an IR excess. Also
\citet{decin2000A&A...357..533D} found 5 Vega-like objects in a
sample of 30 G dwarfs with no bias with respect to age and the
very occurrence of an infrared excess. These fractions are in good
agreement with the fraction of main-sequence stars displaying the
Vega phenomenon as obtained by IRAS, namely $13\pm 10$\%
\citep{plets1999A&A...343..496P}.

We want however to stress that the evolution of debris disks in a
cluster can be differ from these around field stars because
cluster dynamical interaction between stars may affect
circumstellar disks in a way which is unlikely for stars in a
field.

\section{CONCLUSIONS}

In this paper, the time dependence of debris disks is
re-investigated from IR studies obtained by ISOPHOT. Our
re-evaluation of these data led to some important conclusions:
{\sl(i)} a large spread of fractional luminosity occurs, but
{\sl(ii)} there are few very young stars with intermediate or 
small IR excesses;
{\sl(iii)} there appears to be a clear maximum for the fractional
luminosity at a given age around \fdust$\approx 10^{-3}$, and
{\sl(iv)} a debris disk is more common around young stars than
around old ones. However, it is clear that this field of research
is eagerly awaiting the possibilities offered by the SIRTF
satellite. The higher sensitivity of this instrument will alow
more accurate measurements of statistically relevant samples in
clusters and should also enable us to find out whether the
scarcity of young stars with real but low excesses is genuine. To
reach these goals a good absolute calibration of the instruments
and significant work on understanding the photospheres of the
individual stars will be required. For studies of the Vega
phenomenon in the field, it will remain important to complement
SIRTF observations with ground-based studies which allow to
constrain the stellar ages to within useful limits.


\acknowledgements
The ISOPHOT data presented in this paper was reduced using PIA, which
is a joint development by the ESA Astrophysics Division and the ISOPHOT
consortium, with the collaboration of the Infrared Analysis and Processing
Center (IPAC) and the Instituto de Astrofísica de Canarias (IAC).
GD is supported by project IUAP P4/05 financed by the
Belgian Federal Scientific Services (DWTC/SSTC).


\clearpage

\begin{figure}
\plottwo{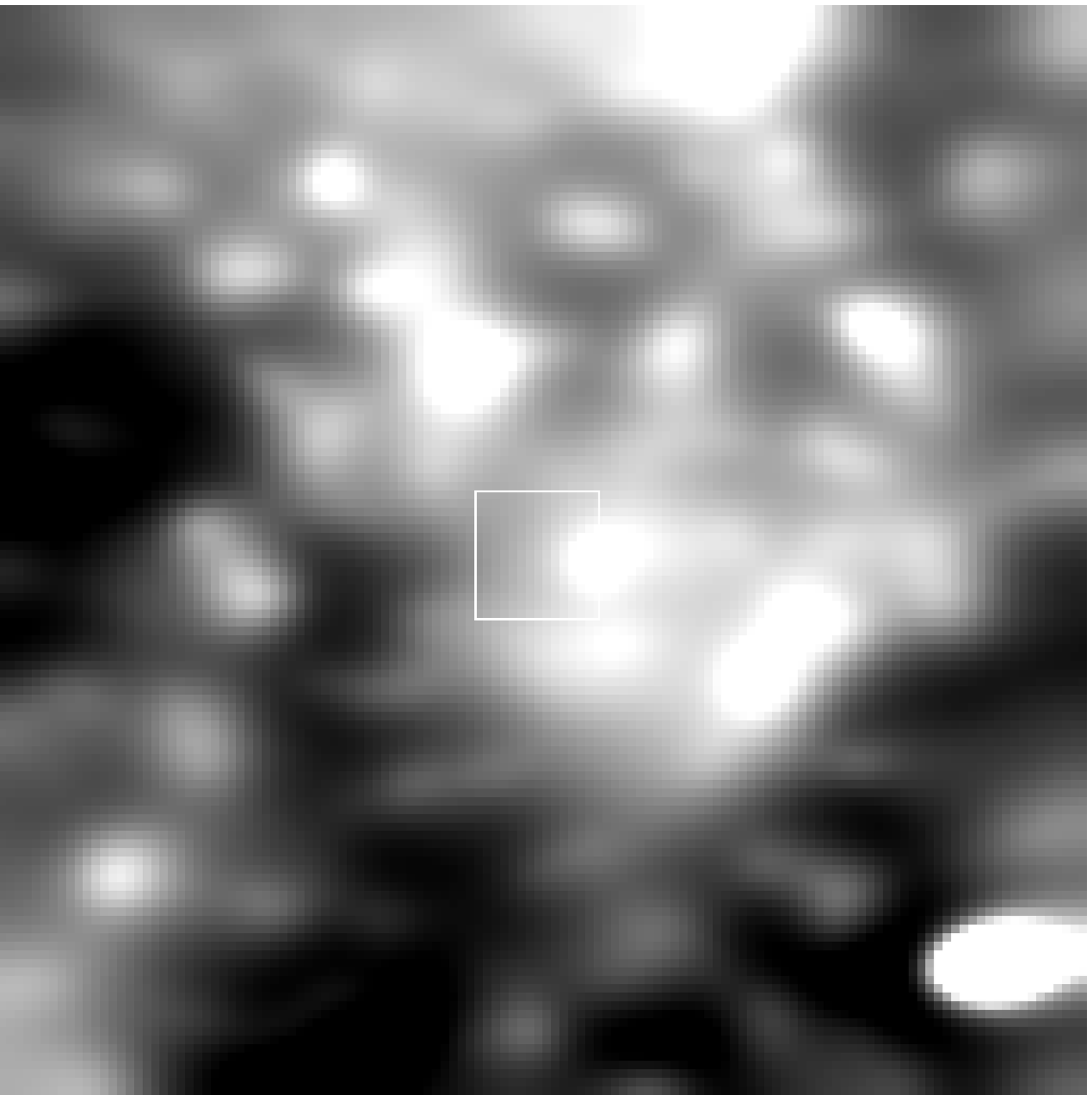}{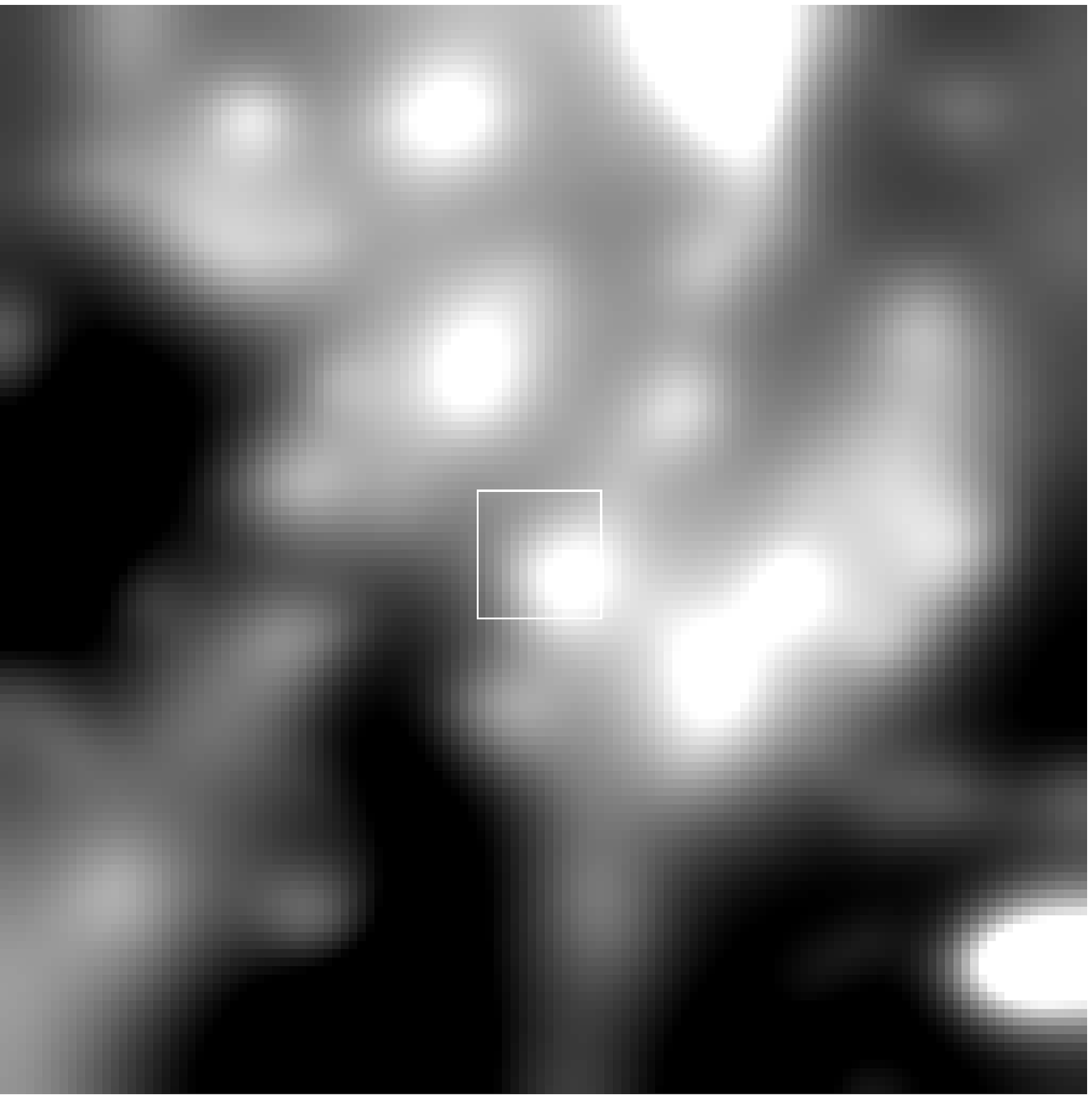} \caption{ The HIRAS maps of the region
around HD35850 at respectively 60~\ums (left) and 100~\ums (right
). The size of each map is 30'$\times$30'. The rectangle, in the
middle of each map, is an indication of the total size of the
region measured by the ISOPHOT C100-raster. \label{iras_hd35850}}
\end{figure}

\clearpage

\clearpage

\begin{deluxetable}{lcrrlccccc}
\tabletypesize{\scriptsize} \tablecaption{Stellar parameters for
stars whose IR excess is most probably due to a debris disk.
\label{data_disk_evolution}} \tablewidth{0pt} \tablehead{
\colhead{name} & \colhead{V} & \colhead{B-V} & \colhead{plx} &
\colhead{Sp Type} & \colhead{$\log{\rm{age}}_{\rm published}$} &
\colhead{$\log{\rm{f_d}}$} & \colhead{$\log{\rm{age}}_{\rm
revised}$} & \colhead{Age} &
\colhead{Reliability}\\
\colhead{}&\colhead{(mag)}&\colhead{(mag)}&\colhead{(mas)}&
\colhead{}&\colhead{(yr)}&\colhead{}&\colhead{(yr)}&\colhead{Reference}&
\colhead{Indicator}} \startdata
\objectname{HD10647}\tablenotemark{1}  & 5.52&  0.551&  57.63& F8V        & 9.54& -3.27&9.54&1&1  \\
\objectname{HD20794}\tablenotemark{1}  & 4.26&  0.711& 165.02& G8V        & 9.86& -4.70&9.86&1&1   \\
\objectname{HD22484}\tablenotemark{1}  & 4.29&  0.575&  72.89& F9V        & 9.72& -4.93&9.72&1&1  \\
\objectname{HD41700}\tablenotemark{1}  & 6.35&  0.517&  37.46& G0IV-V     & 9.63& -3.75&9.63&1&1  \\
\objectname{HD53143}\tablenotemark{1}  & 6.81&  0.786&  54.33& K0IV-V     & 8.99& -3.25&8.99&1&1  \\
\objectname{HD10700}\tablenotemark{2}  & 3.49&  0.727& 274.17& G8V        & 9.86& -4.60&9.86&2&2  \\
\objectname{HD17925}\tablenotemark{2}  & 6.05&  0.862&  96.33& K1V        & 7.90& -3.90&7.90&2&2  \\
\objectname{HD22049}\tablenotemark{2}  & 3.72&  0.880& 311.00& K2V        & 8.51& -3.54&8.51&2&1  \\
\objectname{HD30495}\tablenotemark{2}  & 5.49&  0.632&  75.10& G3V        & 8.32& -4.10&8.32&2&1  \\
\objectname{HD38678}\tablenotemark{2}  & 3.55&  0.104&  46.47& A2Vann     & 8.57& -4.70&8.57&2&1  \\
\objectname{HD95418}\tablenotemark{2}  & 2.34&  0.033&  41.07& A1V        & 8.56& -5.00&8.56&2&1  \\
\objectname{HD102647}\tablenotemark{2} & 2.14&  0.090&  90.16& A3Vvar     & 8.38& -4.80&8.38&2&1  \\
\objectname{HD128167}\tablenotemark{2} & 4.47&  0.364&  64.66& F3Vwvar    & 9.23& -5.00&9.23&2&1  \\
\objectname{HD139664}\tablenotemark{2} & 4.64&  0.413&  57.09& F5IV-V     & 9.05& -4.00&9.05&2&1  \\
\objectname{HD172167}\tablenotemark{2} & 0.03& -0.001& 128.93& A0Vvar     & 8.54& -4.80&8.54&2&1  \\
\objectname{HD207129}\tablenotemark{2} & 5.57&  0.601&  63.95& G2V        & 9.78& -3.80&7.60&6&4  \\
\objectname{HD216956}\tablenotemark{2} & 1.17&  0.145& 130.08& A3V        & 8.34& -4.30&8.34&2&1  \\
\objectname{$\beta$~Pic}\tablenotemark{2} & 3.85&  0.170 &  51.90& A3V        & 8.45& -2.82&7.08&7&4 \\
\objectname{HE 361}\tablenotemark{3}   & 9.68&  0.430&\nodata& F4V        & 7.70& -3.21&7.70&3&4  \\
\objectname{HII 1132}\tablenotemark{3} & 9.43&  0.500&\nodata& F5V        & 8.07& -3.37\tablenotemark{11}&8.07&3&4 \\
\objectname{HD72905}\tablenotemark{3}  & 5.63&  0.618&  70.07& G1.5Vb     & 8.48& -4.69\tablenotemark{11}&8.48&3&4 \\
\objectname{HD125451}\tablenotemark{3} & 5.41&  0.385&  38.33& F5IV       & 8.48& -4.11&8.48&3&4 \\
\objectname{HD139798}\tablenotemark{3} & 5.76&  0.353&  27.98& F2V        & 8.48& -4.48&8.48&3&4  \\
\objectname{HD107067}\tablenotemark{3} & 8.69&  0.523&  14.54& F8         & 8.70& -2.85&8.70&3&4  \\
\objectname{HD108102}\tablenotemark{3} & 8.12&  0.534&   9.34& F8         & 8.70& -3.23&8.70&3&4   \\
\objectname{HD108651}\tablenotemark{3} & 6.63&  0.212&  12.66& A0p        & 8.70& -3.89&8.70&3&4 \\
\objectname{HD105}\tablenotemark{3,4}    & 7.51&  0.595&  24.85& G0V        & 8.67& -3.37&7.93&8&4 \\
\objectname{HD35850}\tablenotemark{3,4}  & 6.30&  0.553&  37.26& F7V:       & 8.37& -4.52&7.08&9&1 \\
\objectname{HD134319}\tablenotemark{3,4} & 8.40&  0.677&  22.59& G5         & 8.70& -3.48&8.78&8&4 \\
\objectname{HD151044}\tablenotemark{3,4} & 6.47&  0.503&  34.00& F8V        & 9.10& -4.17&9.68&11&1 \\
\objectname{HD202917}\tablenotemark{3,4} & 8.65&  0.690&  21.81& G5V        & 8.20& -3.33&7.93&8&4 \\
\objectname{HD209253}\tablenotemark{3,4} & 6.63&  0.504&  33.25& F6/F7V     & 8.60& -3.82&8.65&11&2 \\
\objectname{HD4614}\tablenotemark{4}   & 3.46&  0.587& 167.99& G0V SB     & 8.60& -5.43&9.60&10&3 \\
\objectname{HD131156}\tablenotemark{4} & 4.54&  0.720& 149.26& G8V+K4V    & 8.90& -5.02&9.30&10&3  \\
\objectname{HD144284}\tablenotemark{4} & 4.01&  0.528&  47.79& F8IV-V     & 9.10& -5.57&9.49&11&2  \\
\objectname{HD152391}\tablenotemark{4} & 6.65&  0.749&  59.04& G8V        & 9.00& -4.30&9.00&4&1  \\
\objectname{HD165341}\tablenotemark{4} & 4.03&  0.860& 196.62& K0V SB     & 9.20& -5.00&9.20&4&1 \\
\objectname{HD8907}\tablenotemark{4}   & 6.66&  0.505&  29.26& F8         & 8.76& -3.62&9.48&11&1 \\
\objectname{HD15115}\tablenotemark{4}  & 6.79&  0.399&  22.33& F2         & 8.70& -3.35&8.70&4&1  \\
\objectname{HD15745}\tablenotemark{4}  & 7.47&  0.360&  15.70& F0         & 8.90& -2.89&8.90&4&2  \\
\objectname{HD22128}\tablenotemark{4}  & 7.59&  0.378&   7.06& A5         & 8.50& -3.18&9.14&11&1 \\
\objectname{HD25457}\tablenotemark{4}  & 5.38&  0.516&  52.00& F5V        & 8.70& -4.07&7.93&8&4 \\
\objectname{HD38207}\tablenotemark{4}  & 8.47&  0.360&\nodata& F2V        & 8.60& -3.00&8.60&4&1 \\
\objectname{HD164249}\tablenotemark{4} & 7.01&  0.458&  21.34& F5V        &\nodata& -2.89&7.08&9&4 \\
\objectname{HD206893}\tablenotemark{4} & 6.69&  0.439&  25.70& F5V        & 8.90& -3.66&8.90&11&2 \\
\objectname{HD221853}\tablenotemark{4} & 7.35&  0.405&  14.04& F0         & 8.90& -3.32&9.25&11&2 \\
\objectname{HD177817}\tablenotemark{4} & 6.00& -0.025&   3.65& B7V        & 8.00& -4.19&8.40&11&2 \\
\objectname{HD3627}\tablenotemark{5} & 3.27&  1.268&  32.19& K3III   & 9.50& -4.46&9.50&5&1 \\
\enddata
\tablerefs{ (1) \citet{decin2000A&A...357..533D}; (2)
\citet{habing2001A&A...365..545H}; (3)
\citet{spangler2001ApJ...555..932S}; (4)
\citet{silverstone2000PhDT........17S}; (5) Decin et al. (2003, in
preparation); (6) \citet{zuckerman2000ApJ...535..959Z}; (7)
\citet{barrado1999ApJ...520L.123B}; (8)
\citet{montes2001MNRAS.328...45M}; (9)
\citet{zuckerman2001ApJ...562L..87Z}; (10)
\citet{fernandes1998A&A...338..455F}; (11) this work.}

\end{deluxetable}

\end{document}